\documentclass{article}
\usepackage{graphicx}
\begin{document} 
\title{{\bf A comparison of extremal optimization with flat-histogram dynamics for finding spin-glass ground states}} 
\author{Jian-Sheng Wang$^{1}$  and Yutaka Okabe$^{2}$
\\
{\small $^1$Singapore-MIT Alliance and Department of Computational Science,}\\ 
{\small National University of Singapore, Singapore 119260, Republic of Singapore}\\
{\small $^2$Department of Physics, Tokyo Metropolitan University,}\\
{\small Hachioji, Tokyo 192-0397, Japan}
}
 
\date{3 October 2002}

\maketitle            

\begin{abstract}
We compare the performance of extremal optimization (EO),
flat-histogram and equal-hit algorithms for finding spin-glass ground
states.  The first-passage-times to a ground state are computed.  At
optimal parameter of $\tau=1.15$, EO outperforms other methods for small 
system sizes, but equal-hit algorithm is competitive to EO, particularly for
large systems.  Flat-histogram and equal-hit algorithms offer
additional advantage that they can be used for equilibrium
thermodynamic calculations.  We also propose a method to turn EO into
a useful algorithm for equilibrium calculations.

Keywords: extremal optimization. flat-histogram algorithm, equal-hit 
algorithm, spin-glass model, ground state.
\end{abstract}

\section{Introduction} 

Optimization with methods motivated from real physical processes is an
active field of research.  Simulated annealing \cite{kirkpatrick} and
genetic algorithm \cite{gen} are two well-known examples.  In
particular, there have been a large variety of methods proposed to
find spin-glass ground states 
\cite{sg-kawashima,sg-dittes,sg-pal,sg-hartmann,sg-chen,sg-berg,sg-houdayer,dall}.
Recently, Boettcher
and Percus \cite{boettcher-percus1,boettcher-percus2} introduced `extremal optimization'
(EO) inspired by models of self-organized criticality \cite{bak},
which gave impressive performance.

Most of the heuristic optimization methods (including simulated
annealing and genetic algorithm) are designed to find ground states
only, thus it is not possible to give correct thermodynamics from a
simulation.  On the other hand, multi-canonical ensemble simulation
\cite{berg}, $1/k$-sampling \cite{stichcombe}, parallel 
tempering \cite{hukushima-nemoto}, and recent
flat-histogram dynamics \cite{wangEPJ} are constructed for equilibrium
thermodynamics, but can also be used as methods for optimization.  A
study of optimization by flat-histogram algorithm on the
two-dimensional spin-glass model is carried out in ref.~\cite{zhen}.
Unlike simulated annealing and other heuristic methods, we note that these
methods do not have any adjustable parameters.  It is useful to know
the efficiencies of this second class of methods when used as an
optimization tool.

In this paper, we make a comparative study of the extremal
optimization and flat-histogram/equal-hit dynamics.  We compare
four algorithms: EO at $\tau=1$ with a continuous approximation in the
probability of choosing a site, original EO at optimal value of
$\tau=1.15$, single-spin-flip flat-histogram dynamics, and equal-hit
algorithm with $N$-fold way.  It is found that EO at the value
$\tau=1.15$ is very good for both two- and three-dimensional Ising
spin glasses.  The equal-hit algorithm with $N$-fold way is also competitive.
For large systems, equal-hit appears even slightly better than EO.
It is useful to have the efficiency of EO but still
give equilibrium results.  To this end, we
introduce a rejection step in EO, thus
turning EO into an equilibrium simulation method.

\section{Single-spin-flip algorithms}

In the following, we specialize our discussion in the context of spin
models, and particularly the spin-glass model
\cite{spin-glass-review}.  The energy function is defined by
\begin{equation}
  E(\sigma) = - \sum_{\langle i,j\rangle} J_{ij} \sigma_i \sigma_j,
\end{equation}
where the spin $\sigma_i$ takes on value $+1$ or $-1$ with $i$ varying
over a hypercubic lattice in $d$ dimensions.  The coupling constant
$J_{ij}$ for each nearest neighbor pair $\langle i, j\rangle$ takes on
a random value of $+J$ and $-J$ with equal probability.  
We impose a constraint $\sum_{\langle i,j\rangle} J_{ij} = 0$.
The spin
glass is known to be one of the hardest problems \cite{NP} to find the
state $\sigma$ that minimizes $E(\sigma)$.

A single-spin-flip with rejection is described by a Markov chain
transition matrix of the form
\begin{equation}
   W(\sigma \to \sigma') = \delta_N(\sigma, \sigma') { 1 \over N} 
   a(\sigma \to \sigma'), \qquad \sigma \neq \sigma',
\end{equation}
where $\delta_N(\sigma, \sigma') = 1$ if $\sigma'$ is obtained from
$\sigma$ by a single spin flip, and 0 otherwise.  The factor $1/N$
represents the random selection of a spin, where $N$ ($=L^d$) is 
the number of spins.  $a(\sigma \to \sigma')$ is the flip rate.
If we choose $a(\sigma \to \sigma')$ according to Metropolis rate,
\begin{equation}
  a(\sigma \to \sigma') = 
            \min\left(1, { f\bigl(E(\sigma')\bigr) 
                     \over f\bigl(E(\sigma)\bigr)} \right),
\end{equation}
we can realize equilibrium distribution with the probability of states
distributed according to $f\bigl(E(\sigma)\bigr)$.  Some choices are:
\begin{equation}
  f(E) = \cases{ \exp\bigl(-E/(k_B T)\bigr),  &  canonical ensemble; \cr
         1/n(E),               & multicanonical ensemble; \cr
         1/ \int_{-\infty}^E n(E') dE', &  $1/k$ sampling, \cr}
\end{equation}
where $T$ is temperature, $k_B$ is Boltzmann constant, and 
$n(E)$ is density of states at energy $E$.

Arbitrary choice of the flip rate $a(\sigma \to \sigma')$ in general 
would not give one
important property of the equilibrium systems, i.e., the microcanonical
property that the probability distribution of the configuration
$\sigma$ is a function of energy $E$ only.  For example, the original
broad histogram rate \cite{oliveira}
\begin{equation}
  a(\sigma \to \sigma') = \min\left(1, { N_{Z-k}(\sigma) \over N_{k}(\sigma) 
      } \right),
\end{equation}
and the random walk rate of Berg \cite{berg-nature} do not have
microcanonical property, where $N_k(\sigma)$ is the number of possible
moves of class $k$ in the state $\sigma$; we associate a class for 
each site $i$ with a number from 0 to
$Z$ ($=2d$) by a scaled energy change $k = \bigl((E(\sigma') - 
E(\sigma))/J +2Z\bigr)/4$, i.e.,
\begin{equation}
  k = { 1\over 2} \sum_{j} (J_{ij}\sigma_i \sigma_j + 1).
\end{equation}
$N_k(\sigma)$ is the number of such sites having a class $k$.

The single-spin-flip version with rejection can be turned into a
rejection-free $N$-fold way \cite{bortz} simulation where a class is
chosen with probability
\begin{equation}
 P_k = { a_k N_k(\sigma)  \over  A(\sigma) }, \quad 
    A(\sigma) = \sum_{k=0}^Z a_{k} N_{k}(\sigma) ,
\end{equation}
where $a_k$ is $a(\sigma \to \sigma')$ associated with an energy
change indicated by class $k$.
A spin in that class is picked up at random, and the flip is always
accepted.  Thermodynamic quantities need to be weighted by a factor
$1/A(\sigma)$.

EO \cite{boettcher-percus1,boettcher-percus2} is somewhat related to $N$-fold way in the sense that a class 
is chosen with some probability $P_k$, and a spin in that class
is picked up and always flipped.  The EO algorithm
can be stated as follows: we classify the site by its `fitness' $k$.
There are $Z+1$ possible values for $k$. In the general EO,
the sites are sorted according to the fitness $k$.  Since there are
only a small number of possible values in the $\pm J$ spin-glass model, the
sorting is not necessary.  We simply make a list of sites in each
category.  We pick a class according to the probability $P_k$, and
then choose a spin in that class and flip with probability one.  The
corresponding transition matrix is then
\begin{equation}
   W(\sigma \to \sigma') = \delta_N(\sigma, \sigma') P_k { 1\over N_k},
   \qquad \sigma \neq \sigma'.
\end{equation}
The original choice of EO is to take
\begin{equation}
 P_k \propto \sum_{ N_0 + N_1 + \cdots + N_{k-1} < i \le 
                    N_0 + N_1 + \cdots + N_{k} } i^{-\tau}, 
 \label{EO-original}
\end{equation}
with $\tau$ being a parameter of the algorithm.  We define the
standard EO to be a continuous approximation to the 
above sum at $\tau=1$ with the
analytical expression by
\begin{equation}
   P_k = { 1 \over \ln(N+1) } \ln { 1+ \sum_{j=0}^k N_j \over
           1 + \sum_{j=0}^{k-1} N_j }.
\end{equation}
The number 1 in the numerator and denominator are introduced somewhat
arbitrarily to avoid divergence when the sum of $N_k$ is zero.
An optimized EO will be the discrete version, Eq.~(\ref{EO-original}),
with $\tau$ that gives best performance; we use $\tau = 1.15$ as
recommended in ref \cite{boettcher-percus2}.  To realize the power-law
distribution, we generate an integer $i = \lfloor \xi^{1/(1-\tau)} \rfloor$,
where $\xi$ is a uniformly distributed random number between 0 and 1,
and pick a corresponding site $i$ ordered by the class.
We have used $\lfloor \cdots \rfloor$ for the floor function.

\section{Comparison of EO with flat-histogram and equal-hit algorithms}

The flat-histogram algorithm \cite{wangEPJ,wang-swendsen-JSP} is a
special choice of the flip rate
\begin{equation}
   a^{{\rm FH}}(\sigma \to \sigma') = \min\left(1, {\langle N_{Z-k} (\sigma')\rangle_{E'} 
                    \over \langle N_k(\sigma) \rangle_E } \right),
\end{equation}
where the angular brackets with subscript $E$ denote a microcanonical
average of the quantity $N_k(\sigma)$ at energy $E$; 
the starting state $\sigma$ has
energy $E$ and the final state $\sigma'$ has energy $E'$.  This
particular choice of the rate gives a flat distribution for the energy
histogram, $H(E) = n(E)f(E) = {\rm const}$, $n(E)$ is density of
states. This is one way to realize multicanonical ensemble.

In the $N$-fold way equal-hit algorithm 
\cite{wang-swendsen-JSP,equalhit}, we perform the usual $N$-fold-way
move (thus rejection-free) which is constructed from the following
single-spin-flip rate:
\begin{equation}
   a^{{\rm EQ}}(\sigma \to \sigma') = 
             \min\left(1, { \langle A \rangle_E \langle N_{Z-k} 
         (\sigma')\rangle_{E'} 
         \over  \langle A \rangle_{E'} \langle N_k(\sigma) \rangle_E } \right).
\end{equation}
We note that $\langle A \rangle_E^{-1} = \langle 1/A \rangle_N$, where
$\langle \cdots \rangle_N$ is average over the $N$-fold way samples.
In equal-hit algorithm, it is guaranteed by construction that
we change state in every move, and the distribution of the visits to 
different energies is flat.

Since the microcanonical averages used in the flip rates are not known
before the simulation,  we use running average to replace the exact
microcanonical average.  It appears that this is a valid approximation
and should converge to the correct values for sufficiently long runs.
For a truly exact algorithm (in the sense of realizing microcanonical
property), it is sufficient with a two-pass simulation.  The first
pass uses a running average; in the second pass, we use a
multicanonical rate determined from the first pass.

\begin{figure}[t]
\includegraphics[width=0.85\columnwidth]{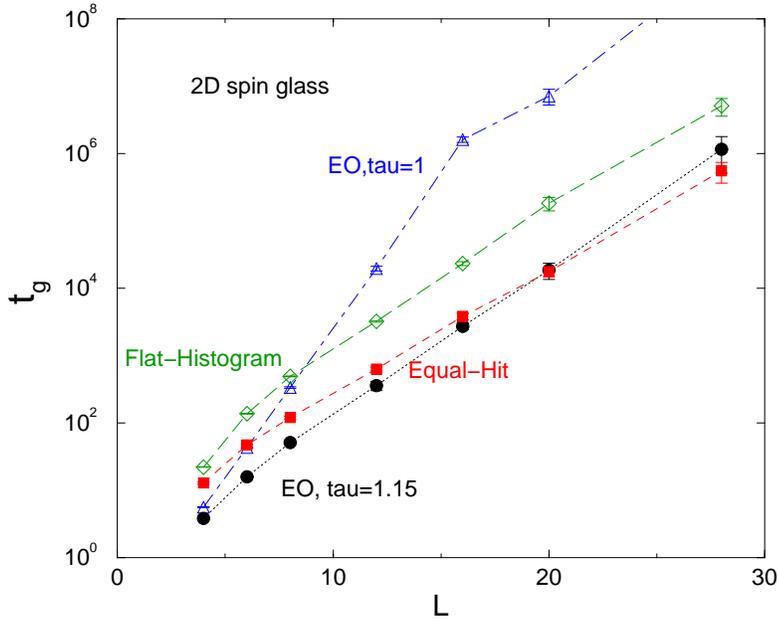}
\caption[tg vs L]{The average first-passage time $t_g$ in units of sweeps 
to find a ground state for four algorithms: standard EO (triangles), 
optimal EO (circles), flat-histogram (diamonds), and equal-hit (squares) 
for the two-dimensional spin-glass model.  Over
$10^3$ realization of random coupling samples are used for averaging 
for each algorithm and size.}
\label{fig-1}
\end{figure}

Many different criteria are used to measure the effectiveness of
an optimization algorithm, such as the fraction of cases for which
ground states are found in a set of runs.  The first-passage-time,
the time in units of Monte Carlo sweeps that a ground state is 
found, starting with similar random configurations, should be
a good measure of the algorithms' efficiency.  We consider sample 
average of the first-passage-time, although the distribution
of it is also very useful.  The computer CPU time is another
useful criterion when comparing algorithms of very different types.

In the flat-histogram and equal-hit algorithms, we can sample positive
as well as negative energies uniformly.  In this study, we have
restricted to the negative energy part, where moves to $E>0$ region are
rejected.  
We compute the average time (first-passage time) for each
lattice size and given algorithm in units of sweeps ($N=L^d$ basic
moves) to find a ground state, starting from a random configuration of
equal probability of spin up and spin down.  
For two-dimensional $\pm J$ spin glass, we determine the first-passage
time $t_g$ by comparing the current value of energy with the exact value 
of ground state energy, obtained from the Spin Glass Server 
\cite{server}.  Thus the results are unbiased.
%%%
The average first-passage time $t_g$ for the two-dimensional Ising spin-glass 
model is shown in Fig.~\ref{fig-1}.  Over 
$10^3$ realization of random coupling samples are used for averaging 
for each algorithm and size.
%%%

\begin{figure}[tb]
\includegraphics[width=0.85\columnwidth]{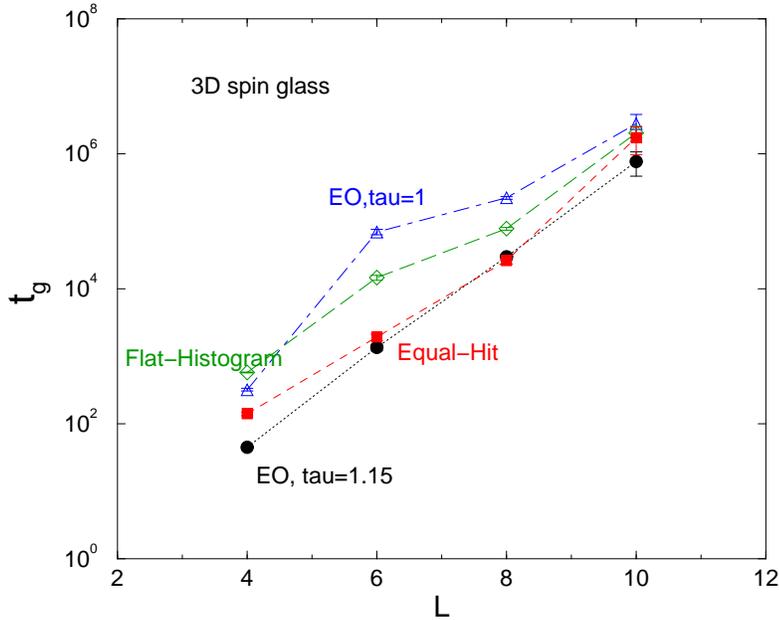}
\caption[tg vs L]{The average limiting time $t_g$ to find a 
ground state for the three-dimensional Ising spin-glass model.  
The meanings of the symbols are the same as those of Figure 1.}
\label{fig-2}
\end{figure}

Since the ground state energies are usually not known in three dimensions, 
we consider instead the time for
finding the lowest energy for a fixed amount of sweeps $t$, averaged
over the coupling constants with the constraint $\sum J_{ij} = 0$.
For any fixed running length $t$, results obtained are only a lower
bound for $t_g$.  We  consider run lengths of $10^4$, $10^5$,
$10^6$, etc, until the first-passage time converges for large $t$.  
This limiting time $t_g$ is reported for the three-dimensional Ising 
spin-glass model in Fig.~\ref{fig-2}.

To compare the efficiency of the four algorithms, the actual CPU times
are also an important factor.  For our implementation, it turns out that the
optimized EO, standard EO, and $N$-fold way equal-hit all have about the
same speed at 6 microsecond per spin flip on a 700 MHz Pentium, while
the single-spin-flip flat-histogram algorithm takes 3 microsecond.
There are several important features in this comparison, 
see Fig.~\ref{fig-1}.  All of them
have a first-passage time that grows rapidly with sizes.  With the
exception of the standard EO, they nearly have the same slope 
of about 6 on a
double logarithmic scale.  It is also interesting to compare the
first-passage time with that of equilibrium tunneling time reported in
ref.~\cite{zhen,wang-swendsen-JSP}.  EO gives excellent performance
for small to moderate size systems.  However, for large sizes,
equal-hit is as good as EO, or even better.  Flat-histogram is worse by
some constant factor.  On the other hand, the performance of the
standard EO at $\tau=1$ is rather poor.  This shows that the results
of EO is rather sensitive to the value of $\tau$.

Another very interesting aspect of Fig.~\ref{fig-1} is that the curves
all look linear in the semi-logarithmic scale.  This implies that
$t_g \sim \exp(c L)$ for some constant $c \sim 1$, not a power law
in $L$.  Thus, all of the algorithms are asymptotically inefficient.
It would be very interesting if this numerical observation
can be supported by some argument. Similar results for the 
three-dimensional Ising spin glass is presented in Fig.~\ref{fig-2}.

\begin{table}
\begin{tabular}{|l|l|l|}
\hline
\multicolumn{3}{|l|}{$L=6$,  MCS $=10^6$,   sample $=1024$} \\
\hline
          & $t_g$         &  $E_g$ \\
EO &             $(6.95 \pm 0.65) \times 10^4$ & $-1.7713 \pm 0.0012$ \\
optimal EO &     $(1.36 \pm 0.15) \times 10^3$ & $-1.7715 \pm 0.0006$ \\
Flat-Histogram & $(1.46 \pm 0.13) \times 10^4$ & $-1.7715 \pm 0.0006$ \\
Equal-Hit  &     $(1.92 \pm 0.20) \times 10^3$ & $-1.7716 \pm 0.0006$ \\
\hline
\multicolumn{3}{|l|}{$L=10$,  MCS $=10^6$,   sample $=256$} \\
\hline
        & $t_g$         &  $E_g$ \\
EO &             $(2.98 \pm 0.17) \times 10^5$ & $-1.7721 \pm 0.0008$ \\
optimal EO &     $(1.04 \pm 0.11) \times 10^5$ & $-1.7809 \pm 0.0008$ \\
Flat-Histogram & $(1.93 \pm 0.15) \times 10^5$ & $-1.7802 \pm 0.0007$ \\
Equal-Hit  &     $(1.27 \pm 0.17) \times 10^5$ & $-1.7815 \pm 0.0010$ \\
\hline
\end{tabular}
\caption[tg vs L]{The average time $t_g$ and lowest energy $E_g$ obtained 
for three-dimensional Ising spin-glass.}
\label{tb1}
\end{table}

In Table~\ref{tb1}, we report some typical data for
the average first-passage time 
$t_g$, energy per site, length of the run, and number of
random samples for four algorithms for the three-dimensional 
spin glasses.  Since we use the same set of samples with the
four algorithms, a lower energy indicates a better performance.
The data show that equal-hit is comparable to EO at optimal
$\tau$.

\section{Turning EO into an equilibrium algorithm}

The flat-histogram and equal-hit algorithms can be used for
equilibrium simulation.  With the help of counting the number of
potential moves, $N_k$, a
basic requirement for obtaining equilibrium property of the simulated
model is the microcanonical property.  Using the broad histogram
equation \cite{oliveira2,berg-hansmann},
\begin{equation}
     n(E) \langle N_k(\sigma) \rangle_E 
  =  n(E') \langle N_{Z-k}(\sigma') \rangle_{E'},
   \qquad k = \bigl( (E'-E)/J + 2Z\bigr)/4,
  \label{broad-histogram-eq}
\end{equation}
we can obtain the density of states $n(E)$ of energy $E$, thus the
equilibrium thermodynamic quantities, including free energy.

Unfortunately, the microcanonical property that the probability
distribution $P(\sigma)$ of the configurations is a function of energy
$E$ only is strongly violated in EO.  The probabilities of the ground
states cluster into groups, rather than uniformly distributed.
Numerical tests show that $ P(\sigma)$ is a function of $E$, $N_k$, as
well as additional unknown parameters.  To correct this problem, we
introduce a rejection step in the EO algorithm, as follows:
\begin{equation}
   W(\sigma \to \sigma') = \delta_N(\sigma, \sigma') 
                         P_k { 1\over N_k} a(\sigma \to \sigma').
\end{equation}
The acceptance rate $a$ is determined by imposing a detailed balance
with an unknown probability distribution $f\bigl(E(\sigma)\bigr)$,
\begin{equation}
  f\bigl(E(\sigma)\bigr) W(\sigma \to \sigma') = 
  f\bigl(E(\sigma')\bigr) W(\sigma' \to \sigma). 
\end{equation}
This gives an equation for the rate $a$:
\begin{equation}
  f(E) P_k { 1\over N_k(\sigma)} a(\sigma \to \sigma') = 
  f(E') P'_{Z-k} { 1\over N_{Z-k}(\sigma') } a(\sigma' \to \sigma). 
\end{equation}
The prime on $P'$ indicates that it is a set of $P$ values calculated
from the state $\sigma'$.  A solution to this equation is a
Metropolis-type choice:
\begin{equation}
  a(\sigma \to \sigma') = \min\left(1, 
  { f(E') P'_{Z-k} / N_{Z-k}(\sigma') \over 
  f(E) P_k / N_k(\sigma) } \right).
\end{equation}

To implement this, we need a two-pass simulation.  The first pass
determines the function $f(E)$.  The procedure is by no means unique.
Here, we collect histogram of energy $H(E)$ as well as statistics for
$\langle N_k \rangle_E$ from an incorrect simulate of the original EO.
Then we determine an approximate density of states with the help of
the broad histogram equation, Eq.~(\ref{broad-histogram-eq}).  The
function $f$ is computed from $f(E) = H(E)/n(E)$.  The EO with
rejection is implemented in the second pass.
The above procedure should be applicable for any model and any
optimization algorithms that has a `steady state'.   If the 
microcanonical property is only slightly violated, it will give
a correct equilibrium algorithm with $a$ nearly equal to 1.
Thus, we hope to have a method that is efficient for optimization,
and yet at the same time, give correct equilibrium statistics.

Indeed, with the above method 
the microcanonical property is restored.  Due to the rejection
step, the dynamics is slightly changed.  A consequence is that the
histogram in the second pass shifted towards high energy side,
thus the efficiency of the original EO is lost. 

\section{Conclusion}

From this study, we show that equal-hit algorithm is an excellent
candidate for ground state search.  At the same time, it also offers the
possibility for equilibrium calculations, such as the computation of
the ground state entropy.  We also show how optimization algorithms
like EO can be turned into equilibrium algorithms by introducing 
a rejection step.
All the algorithms studied here give rather
rapid increase of $t_g$ with sizes, thus it is important and challenging to
find algorithms that reduce this growth.  Perhaps, algorithms based on
single-spin-flip have their fundamental limitations.

\section*{Acknowledgements}
J.-S. W. thanks the hospitality of Tokyo Metropolitan University
during part of his sabbatical leave stay.  We also thank N. Kawashima
and K. Chen for discussions.  We thank M. Iwamatsu for drawing our attention
to EO algorithm.

\end{document}